\documentclass[fleqn,twoside]{article}
\usepackage{espcrc2}
\usepackage{amsmath}  
\usepackage{verbatim}

\usepackage{graphicx}
\usepackage[figuresright]{rotating}

\input epsf.tex
\newcommand{\unit}[1]{\; \mbox{#1}}

\newcommand{\lsim}{\raisebox{-0.13cm}{~\shortstack{$<$ \\[-0.07cm] $\sim$}}~}

\title{ \bf An indirect dark matter search with diffuse gamma rays from the Galactic Centre with the Alpha Magnetic Spectrometer}

\author{A. Jacholkowska\address[MCSD]{Laboratoire de Physique Th\'eorique et Astroparticules, UMR5207-UM2/IN2P3-CNRS,\\
Place Eug\`ene Bataillon - CC70, 34095 Montpellier, France} \thanks{E-mail: Agnieszka.Jacholkowska@cern.ch},
        G. Lamanna\address{Centre de Physique des Particules de Marseille, UMR/IN2P3-CNRS,\\
                 163 avenue de Luminy - Case 902, 13288 Marseille, France},
        E. Nuss\addressmark[MCSD],
        J. Bolmont\addressmark[MCSD],
        C. Adloff\address[LAPP]{Laboratoire d'Annecy-le-Vieux de Physique des Particules, LAPP/IN2P3-CNRS et Universit\'e de Savoie, F-74941 Annecy-le-Vieux, France},
        J. Alcaraz\address[CIEMAT]{Centro de Investigaciones Energ\`eticas, Medioambientales y Tecnol\`ogicas,\\
        CIEMAT, E-28040 Madrid, Spain},
        R. Battiston\address[INFN]{University and Sezione INFN of Perugia, Italy},\\
        P. Brun\addressmark[LAPP],
        W.J. Burger\addressmark[INFN],
        V. Choutko\address{Laboratory for Nuclear Science, MIT, 77 Massachusetts AV., Cambridge MA 02171-9131, USA}, 
        G.Coignet\addressmark[LAPP],
        A. Falvard\addressmark[MCSD],
        E. Fiandrini\addressmark[INFN],
        L. Girard\addressmark[LAPP],
        C. Goy\addressmark[LAPP],\\
        K. Jedamzik\addressmark[MCSD],
        R. Kossakowski\addressmark[LAPP],
        G. Moultaka\addressmark[MCSD],
        S. Natale\address[DPNC]{DPNC, University of Geneva, 24, Quai Ernest-Ansermet, 1211, Geneva 4, Switzerland },
        J. Pochon\addressmark[LAPP],
        M. Pohl\addressmark[DPNC],
        S. Rosier-Lees\addressmark[LAPP],\\
        M. Sapinski\addressmark[MCSD]\thanks{On leave from Henryk Niewodniczanski Institute of Nuclear Physics in Cracow},
        I. Sevilla Noarbe\addressmark[CIEMAT],
        and JP. Vialle\addressmark[LAPP],}

\begin{document}

\begin{abstract}

The detection of non-baryonic dark matter through its gamma-ray annihilation 
in the centre of our galaxy has been studied. The gamma fluxes according 
to different models have been simulated and compared to those expected to be observed with the
Alpha Magnetic Spectrometer (AMS), during a long-term mission 
on board of the International Space Station. Under the assumption that the dark 
matter is composed of the lightest, stable supersymmetric particle, 
 the neutralino, the results of the simulations in the framework of $mSUGRA$ models, 
show that with a cuspy dark matter 
halo profile or a clumpy halo, the annihilation
gamma-ray signal would be detected by AMS. More 
optimistic perspectives are obtained with the Anomaly Mediated Supersymmetry Breaking
 {\it(AMSB)} model.
The latter leads also to a cosmologically important $^6$Li abundance. Finally, the discovery potential for the massive Kaluza-Klein dark matter candidates has been evaluated and their detection looks feasible. 

\vspace{1pc}
\end{abstract}

\maketitle

\section{Introduction}
The nature of 
dark matter is one of the outstanding questions and challenges in cosmology.
The existence of cosmological dark matter is required by a multitude of 
observations and arguments, such as the excessive peculiar velocities of 
galaxies within clusters of galaxies, or the observations of gravitational arcs  
indicating much deeper gravitational 
potentials within clusters than 
those inferred by the presence of the luminous matter~\cite{BertoneRev52,BertoneRev171}. On the galactic scale, extensive dark matter halos 
are required to explain the observed rotation curves
in spiral galaxies, or the velocity dispersion in
elliptical galaxies~\cite{BertoneRev46,BertoneRev507}. 
Furthermore, Big Bang nucleosynthesis predicts a fractional 
contribution of baryons to the critical density, $\Omega_b$, significantly 
smaller than the total $\Omega_{m}$ in form of clumpy matter.
The Wilkinson Microwave 
Anisotropy Probe (WMAP) has provided the most detailed measurements of the Cosmic 
Microwave Background (CMB) anisotropies~\cite{dona1}. 
In the framework of the Standard Cosmological Model, WMAP quotes a 
total matter density of
  $\Omega_{m}$= 0.27$\pm$0.04  and a baryon density of $\Omega_b$ = 0.044$\pm$0.004, 
which confirms that most of the matter is non-baryonic, in agreement with the results obtained from 
primordial nucleosynthesis studies.

\indent
Various non-baryonic dark matter candidates require physics beyond the standard model 
of particle physics (for a recent review see e.g.~\cite{BertoneRev}). 
N-body simulations of structure formation ~\cite{pri} suggest a non-relativistic, 
Weakly Interacting Massive Particle (WIMP) as a dark matter component, thus favoring
the Cold Dark Matter scenario ~\cite{kamionkowski,bergpr}. 
The WMAP measurement of the density of the non-baryonic dark matter 
provides constraints in the range of 0.095 $<$ $\Omega_{CDM}$$h^2$ $
<$ 0.129, at the 2 $\sigma$ level. 

Supersymmetric theories offer an excellent WIMP candidate, which satisfies the CDM
paradigm and the constraints on $\Omega_{CDM}$, namely the  neutralino 
($\chi_{1}^{0}$) of the Minimal Supersymmetric Standard Model (MSSM), assumed to
be the Lightest 
Supersymmetric Particle (LSP) and stable due to $R$-parity conservation~\cite{goldberg}. At present, 
lower limits on the LSP neutralino mass in the $MSSM$  are about
 50 GeV from LEP experiments (although the mass may be significantly smaller depending on the assumptions relative to gaugino mass universality).
Less conventional scenarios than the neutralino within the minimal supergravity 
(mSUGRA) context have been proposed 
\cite{mSUGRA}:

\begin{itemize}
\item{} In the Anomaly Mediated Supersymmetry
Breaking ($AMSB$) scenario \cite{randal10,randal11} 
the neutralino LSP is predominantly a Wino [the supersymmetric partner of the
electrically neutral component  of the $SU(2)_L$ gauge bosons]. Endowed with a 
relatively large annihilation cross section this particle may constitute the bulk of the dark matter when subsequently to its thermal freeze-out, it is further generated nonthermally (e.g. via Q-ball evaporation or gravitino decay). By virtue of its large annihilation cross section
the wino may lead to possibly large gamma fluxes~\cite{AMSBHooper}.

\item{} In extra-dimension models, ultimately motivated by string theories, 
it has been argued~\cite{kk1} that the lightest
Kaluza-Klein excitation  can provide under certain conditions, 
a very good CDM candidate. In the present paper we will restrict ourselves to
the possibility of low scale extra-dimensions as an extension 
of the non-supersymmetric standard model~\cite{appelquist}
with a perfectly viable dark matter Kaluza-Klein particle~\cite{servant}. 
\end{itemize}

In this paper, we present the predicted $\gamma$-ray fluxes from the Galactic Centre
from neutralino annihilations in the 
frame of $mSUGRA$ and $AMSB$ models, as well as from Kaluza-Klein dark matter annihilations.
The predicted fluxes are used to assess, for the different scenarios,  the discovery potential for non-baryonic dark matter provided by a three-year observation of the diffuse $\gamma$-ray differential spectrum by the AMS on the ISS.

\section{Model descriptions and simulations}

The limited knowledge of dark matter structure and the density profile near the Galactic Center represent the principal astrophysical uncertainties when evaluating the discovery potential of the dark matter through indirect detection. On the other hand, the predicted $\gamma$-ray fluxes depend on the assumptions made within the framework  
 of the particle physics models associated with the different dark matter candidates.

\subsection{Dark matter halo parametrization}

One may parameterize the mass density profile of our Galaxy by the following equation:
\begin{equation}
\rho_{\chi}(r) \; = \; \rho_0 \,
\left( \frac{R_0}{r} \right)^\gamma \,
\left\{ {\displaystyle
\frac{R_0^\alpha + a^\alpha}{r^\alpha + a^\alpha}}
\right\}^\epsilon \;\;,
\label{neutralino_rho}
\end{equation}
assuming a simple spherical Galactic halo.
An isothermal profile with core radius $a$
 corresponds to $\gamma = 0$, $\alpha = 2$ and
$\epsilon = 1$ as proposed by ~\cite{sal}. A Navarro, Frenk and White (NFW)
profile ~\cite{nav} is obtained with $\gamma = 1$, 
$\alpha = 1$, and $\epsilon =2$, whereas Moore's distribution ~\cite{bb6}
 is recovered if $\gamma = \epsilon = 3/2$ and $\alpha = 1$.
Only the NFW and Moore models are considered in this study. The two models predict large values of
the neutralino density in the Galactic Centre (GC).

\noindent
 The parameters of the halo modeling are:
 \begin{itemize}
 \item $R_0$ - distance from Earth to GC,
 \item $\rho_0$ - halo density at $R_0$,
 \item $a$ - the core radius - for $r<a$ the halo density
 is constant and equal to $\rho(a)$ in case of the isothermal
 parametrization.
 \end{itemize}

\noindent
 As shown in ~\cite{bergGal}, $\rho_0$ and $a$ cannot be chosen
arbitrarily. The total mass of
the Galaxy restricts the ($\rho_0$, $a$) parameter space.
For the NFW-$standard$ model the generic parameter values are:
$R_0= 8.0~kpc$, $\rho_0 = 0.3~GeV/cm^3$,  $a = 20~kpc$.
Another possible combinations of ($\rho_0$, $a$) parameters allow, given the uncertainties: $R_0= 8.5~kpc$, $\rho_0 = 0.4~GeV/cm^3$,  $a = 4~kpc$ (NFW-$cuspy$).
The two configurations were considered.
 The values for the Moore profile have been chosen as follows:
 $R_0= 8.0~kpc$, $\rho_0 = 0.3~GeV/cm^3$, $a = 28~kpc$.

\noindent

The WIMPs located around the Galactic center should annihilate and 
produce high energy photons. The corresponding photon flux near the Earth, $\Phi_{\gamma}$
-- per unit time, surface, and 
solid angle -- may be expressed as

\begin{equation}
\Phi_{\gamma}\; = \; \frac{1}{4 \pi} \,
{\displaystyle
\frac{\langle \sigma v \rangle \, N_{\gamma}}{2\;m_{wimp}^{2}} } \,
{\displaystyle \int}_{\rm los} \rho_{wimp}^{2}(r) \, ds \;\;. 
\label{gr_flux_1}
\end{equation}

In equation 2, $m_{wimp}$ is the mass of the WIMP-dark-matter candidate; $\langle \sigma v \rangle$ denotes the thermally averaged annihilation rate
; $\rho_{wimp}(r)$ is the mass density of the dark matter and $r$ is the distance from the Galactic center. The flux $\Phi_{\gamma}$ is proportional to the number of annihilations per unit time and volume, $\langle \sigma v \rangle$ $\rho_{wimp}^{2}(r)$/$m_{wimp}^{2}$ and to the number of secondary photons per annihilation, $N_{\gamma}$. Finally to obtain the flux at the Earth it is necessary to integrate the WIMP density squared along the line-of-sight (los) connecting the observer to the Galactic Center. The integral can be expressed in the form:

\begin{equation}
J(R) \; = \; 2 \,
{\displaystyle
\int_{0}^{\sqrt{R_{\rm 0}^{2} - R^{2}}}} \,
\rho^{2} \bigg (\sqrt{s^2 + R^2}\bigg)\; ds \;\;,
\label{gr_flux_2}
\end{equation}
assuming a spherical halo with radial extension $R_{\rm 0}$. The coordinate $s$ extends along the line of sight. $R$ is the radial distance from the centre of the Galaxy for such a direction.

Because the
density decreases steeply at large distances (see Eq.~\ref{neutralino_rho}), a radial cut-off $R_{\rm c}$
 has been applied. It has been checked that
our results are not sensitive to the value
of $R_{\rm c}$, which is set to 8.0 or 8.5 kpc depending on the chosen halo profile.
%


We integrate the function $J$ over a solid angle around the Galactic center, subtended by the detector acceptance, $e.g.$ a circular region with angular radius $\theta_{\rm obs}$:
%
\begin{equation}
\Sigma \; = \; 2 \pi \,
{\displaystyle \int_{0}^{\displaystyle \theta_{\rm obs}}}
\, J(R) \; \sin \theta \; d \theta \;\; ,
\label{le19}
\end{equation}
where $R /R_{0} = \tan \theta \simeq \theta$. 
Thus the resulting value for $I_{\gamma}$ -- flux of high-energy photons
collected per unit of time and surface -- can be written as:

\begin{equation}
\! \! \! \! \!   {\scriptsize I_{\gamma} = 
\left( 3.98 \times 10^{-18} \unit{photons} \unit{cm}^{-2}
\unit{s}^{-1} \right) \times  \nonumber  
}
\end{equation}
\begin{equation}
{\scriptsize
\;\;\;\ \times \left( {\displaystyle
\frac{\langle \sigma v \rangle \, N_{\gamma}} 
{10^{-29} \unit{cm}^3 \unit{s}^{-1}}} \right) \times 
\left( {\displaystyle \frac{1 \unit{TeV}}{m_{\chi}}}
\right)^{2} \; \Sigma_{19} \;\; , 
\label{I_gamma}
}
\end{equation}

\noindent
where $\Sigma_{19}$ denotes $\Sigma$ expressed in units
of ${10^{19} \unit{GeV}^{2} \unit{cm}^{-5}}$ and integrated over the AMS acceptance, $\Delta \Omega$ = 10$^{-3}$ sr. 
We have integrated the relation ~(\ref{le19}) as a function of galactic halo profiles leading to the following results:
\begin{itemize}
\item For a NFW-$standard$ profile : $\Sigma_{19}$ = 2.7 10$^{2}$,  thus $\tilde{J}(0) (\Delta \Omega)$ =   1.2 10$^{3}$ \\
\item For a NFW-$cuspy$ profile: $\Sigma_{19}$ = 117.7 10$^{2}$,  thus $\tilde{J}(0) (\Delta \Omega)$ =  50.0 10$^{3}$\\
\item For a Moore profile: $\Sigma_{19}$ = 336.7 10$^{2}$,  thus $\tilde{J}(0) (\Delta \Omega)$ =  142.9 10$^{3}$\\
\end{itemize}

\noindent
where $\tilde{J}(0) (\Delta \Omega)$ corresponds to the notation used in~\cite{bergGal}. For integration, we have defined an inner cutoff radius at $R_c = 10^{-5}\,$pc such that for $R > R_c$ the flux vanishes.


For completeness, two astrophysical factors may enhance the expected gamma fluxes from
neutralino annihilations:
\begin{itemize}

\item
clumpiness of the dark matter halo as indicated by N-body simulations
 \cite{pri},
\item
the presence of a Supermassive Black Hole (SBH) with a mass of
$\sim 2.6 \times 10^6~M_\odot$ 
creating unstable conditions due to baryon infall by the adiabatic compression process, studied by~\cite{bertone2} 
\end{itemize}
The overall enhancement factor of the expected flux is estimated
in~\cite{M31} to be between 5 and 100, 
depending on the clumpiness of the galactic halo. The enhancement of the annihilation signal in presence of a spike in the dark matter halo is significant with respect to ordinary dark matter cusp, even in case of gravitationnal scattering of stars and the self-annihilating dark matter particles, as pointed by \cite{ullios,edsjo,bertone2}.

\subsection{Models for WIMP candidates}

In this section, we describe briefly the physics models associated with selected dark matter candidates, including the methodology and assumptions used to investigate the different hypotheses.
\subsubsection{mSUGRA parameterization}
The two supersymmetric scenarios considered belong to the class 
of models where supersymmetry (SUSY) breaking is effectively communicated to the 
visible sector via (super)-gravitational effects. 
We use the conventional mSUGRA scenario~\cite{mSUGRA}
 with common values for
the soft supersymmetry breaking scalar and gaugino masses and
trilinear couplings, $m_0, m_{1/2}, A_0$, taken as initial conditions at a 
given high energy universality scale. We require the three gauge couplings to take a common value
at a unification scale $M_{GUT}$ and, for simplicity, identify this scale with the universality
scale of the soft SUSY parameters.       
With these initial conditions and with the value of $\tan \beta$ 
(the ratio of the two Higgs vacuum expectation values
$\frac{<H_2^0>}{<H_1^0>}$) defined at the electroweak scale,  the relevant low 
energy quantities are obtained throug the running of the 
 parameters from $M_{GUT}$ down
to a scale of the order of the electroweak scale.  Electroweak 
symmetry breaking is then required at that scale with the correct Z boson 
physical mass,
thus fixing the supersymmetric $\mu$ parameter (up to a sign)  and its
soft supersymmetry breaking counterpart.

The flux predictions are obtained by use of computational tools 
which allow to scan over various SUSY parameters.
This is achieved through an interface of the two codes,
DarkSUSY~\cite{DSUSY} and SUSPECT\cite{Suspect}, which we dub 
hereafter DSS (DarkSUSY-SUSPECT). Significant features of particle physics 
and cosmology are thus combined in our approach, taking into 
account various phenomenological constraints  
(consistency of the top, bottom and $\tau$ masses, present experimental 
limits on the superpartner and Higgs masses, limits from  $ b \to s \gamma$,
no charged LSP, ..., relic density constraints), some of which are implemented 
in SUSPECT and others in DarkSUSY. 
We have checked in the mSUGRA framework the
compatibility of the results obtained with the DSS software package and the ISASUGRA interface provided with DarkSUSY. 
\subsubsection{Anomaly Mediated Supersymmetry Breaking parameterization}

The Anomaly Mediated Supersymmetry Breaking {\it(AMSB)} is a  gravity-mediated 
mechanism where the SUSY breaking is communicated to the observable sector by the super-Weyl 
anomaly~\cite{randal10,randal11}. In particular, 
the masses of gauginos are generated at the one-loop level as in \cite{randal}:
\begin{equation}
M_i = b_i \left( \frac{\alpha_i}{4\pi} \right)^2 <M>
\end{equation}
where $\alpha_i$ are the gauge coupling constants and $b_i$  the 
associated $\beta$-function coefficients. 
$M$ is the auxiliary field in the supergravity multiplet whose vacuum 
expectation value $<M>$ is expected to be of the order of the gravitino 
mass $m_{3/2}$, the latter being generically in the range: 
10 TeV $<$ $m_{3/2}$ $<$ 100 TeV.

In the minimal AMSB model, the sleptons suffer typically 
from a tachyonic problem. One way to fix this problem is to add a 
scalar mass parameter $m_0^2$, accounting for a non-anomalous contribution
to the soft SUSY breaking.


The most important message from the gaugino mass formula above is the hierarchy:
$$
M1 : M2 : M3 = 2.8 : 1 : 8.3
$$
as opposed to $M1$ : $M2$ : $M3$ = $1$ : $2$ : $7$ which is expected for the gravity- or gauge-mediated models. This implies that the lightest neutralino ( ${\tilde{\chi}_1^0}$) and the lightest chargino ( $\tilde{\chi}_{1}^{\pm}$) are almost pure Winos and consequently mass-degenerate.

\subsubsection{Kaluza-Klein Dark Matter} 
Models with compact extra dimensions predict several new states, the Kaluza-Klein (KK) excitations.
 In the case of Universal Extra Dimensions (UED) \cite{appelquist}, all Standard Model 
fields can propagate in the bulk, and their effective four-dimensional interactions with the KK states
conserve  a quantum number associated with the latter. The conservation of 
this KK number implies that the KK modes  cannot decay exclusively into Standard Model particles; 
the lightest KK mode (LKP) will thus be stable~\cite{appelquist,servant_23}.
Moreover, the LKP, when electrically neutral and with no baryonic charge, provides a viable dark 
matter candidate~\cite{servant}. 
The mass of the LKP dark matter particle, like all other states of the KK tower,
 is inversely proportional to the compactification radius $R$. Accelerator electroweak 
 measurements constrain rather weakly the UED scenario, since in this scenario  the observables 
 are sensitive only to the virtual effects of the KK modes. The lower mass bound leads 
 to $R^{-1}$ $\geq$ 280 GeV~\cite{appelquist}. 
 The most promising LKP dark matter candidate is associated with the first level of KK modes of 
 the Hypercharge gauge boson $B^{(1)}$.
In our calculation we consider the relic density of $B^{(1)}$ of~\cite{servant} leading to a $B^{(1)}$ lower bound mass bound constraint of 
400 GeV.

\section{AMS gamma detection and sensitivity}

\subsection{The AMS-02 experiment}

\begin{figure}[!ht]
\begin{center}
\includegraphics[clip,scale=0.5]{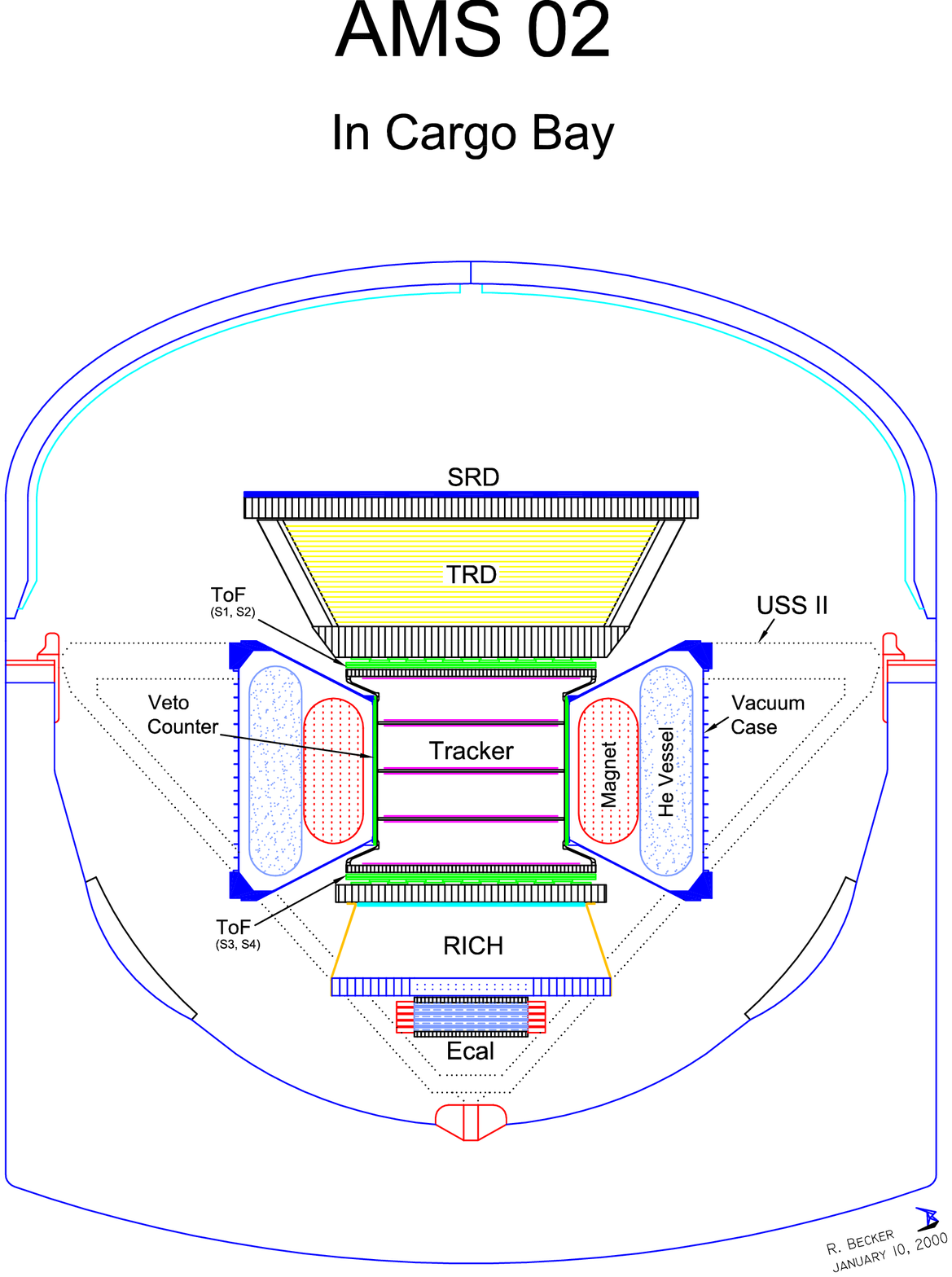}
\end{center}
\caption{Schematic view of the AMS-02 experiment which will operate on the International Space Station.
\label{ams2}}
\end{figure}
The main elements of AMS-02 detector ~\cite{AMS02}
 which
are shown  in Figure~\ref{ams2} include:  
a superconducting magnet, a gaseous transition radiation  detector (TRD), a silicon tracker (Tracker), time-of-flight 
hodoscopes (TOF), a ring imaging Cerenkov  detector (RICH), an electromagnetic calorimeter (ECAL) and anti-coincidence Veto
counters. The superconducting magnet  has the shape of a cylindrical shell
with the inner diameter of 1.2\,m and length of 0.8\,m; it provides a central dipole field of 0.8 Tesla. The eight layers of double-sided silicon tracker sensors are placed in planes  transverse to the magnet axis. 
The silicon tracker measures the
trajectory of relativistic singly charged particles with an
accuracy of 10\,$\mu$
in the bending and 30\,$\mu$ in the non-bending coordinates. 
It provides also measurements of the particle energy loss which allows to distinguish the charge.  
The time-of-flight system (TOF) containing four detection layers, measures
singly charged particle transit times with
an accuracy of 140\,psec and also yields  energy loss and coordinate measurements.
The transition radiation detector (TRD) is situated on the top of the spectrometer and consists of twenty 12\,mm thick foam radiator arrays, interleaved
by arrays of  6\,mm diameter  gas proportional tubes filled with a Xe/CO$_2$ mixture. The TRD provides an $e^-$/hadron separation better than one hundred up to an energy of 200 GeV as well as precise charged particle coordinate measurements.
The RICH detector is installed below the last TOF plane and consists of a 3\,cm thick aerogel radiator with a refraction index of 1.05, a mirror and pixel type matrix photo-tubes for the light detection measures of the velocity of the single charged particle with an accuracy better than 
a fraction of a percent.
The ECAL detector is situated at the bottom of the AMS02 setup. It is a three-dimensional ($65 x 65 x 17 cm^{3}$ electromagnetic sampling calorimeter with total length of 16$X_0$, consisting of 1 $mm$ diameter  scintillating fibers sandwiched between grooved lead plates.

\subsection{Performance of photon detection}




Cosmic $\gamma$-rays may be detected in AMS by two different methods. The {\it conversion mode} involves the 
reconstruction in the tracker of the $e^+e^-$ pairs produced by $\gamma$ conversions in the material upstream of the first layer of silicon sensors \cite{roberto,lamannaelba,lamannanote}.
In the {\it single photon mode}, the $\gamma$-rays are detected in the electromagnetic calorimeter \cite{vitaly}.

The performance of AMS02 detector for $\gamma$-rays has been studied with the AMS simulation and reconstruction program based on GEANT \cite{GEANT}. The simulated performances have been validated using the AMS01 data for the sub-detectors present during the shuttle test flight \cite{alcaraz} and the test beam data obtained with prototypes of the new or modified modules.

The Monte Carlo sample for the present study includes more than $10^9$ reconstructed events including both cosmic $\gamma$-rays and charged-particle backgrounds over the relevant energy range. The latter in order of decreasing importance include protons, He and C nuclei and electrons. With a charged particle background rejection of O(10$^4$) to O(10$^5$), we obtain a background-to-signal ratio of the order of a few percent. The principal background at this level is due to the galactic diffuse  $\gamma$-ray emission.

\subsubsection{ The Conversion Mode  }
The event signature for this mode are two reconstructed tracks in the Tracker
originating from a vertex located upstream of the first silicon layer of the tracker\footnote{The material in front of the first silicon tracker plane, consists of the TRD, the first two layers of ToF scintillators, and mechanical supports, represents $\simeq$ 0.23$X_0$.}.



The incident $\gamma$-ray energy and direction were determined by 
adding the reconstructed momenta components of the $e^\pm$ pair, evaluated at the 
entrance of the AMS02 detector.

%

The main source of background are $p$ and $e^-$ which interact in the AMS detector, producing secondaries, mainly delta rays,
which result in double-track events associated with a common origin at the interaction point.

%

The conversion of the secondary photons
produced in the vicinity of the AMS, i.e. 
in the ISS body and solar panels, was found to be negligible in
comparison to the expected $\gamma$-ray fluxes.

The following criteria are applied to reject background events:

\begin{itemize}
\item Identify events with interactions;
\item Identify charged particles entering the TRD from the top and fire all the tubes along its reconstructed trajectory. 
\item Identify reconstructed large invariant mass events.
\item Identify particles entering the fiducial volume of the AMS through the side of the TRD ).
\end{itemize}

A preliminary rejection factor of $5 \times 10^4$ was obtained for each different cosmic ray species (e$^-$ and $p$), after all selection cuts 
have been applied.

\subsubsection{ The Single Photon Mode  }

The  event signature for this mode is the presence of electromagnetic-type 
energy deposition in the ECAL, while almost nothing is found in the other AMS sub-detectors. 

The identified backgrounds contributing to the cosmic $\gamma$-ray signal are  events with charged particles\footnote{mostly $e^-,~p$ and $He$ nuclei} 
either passing undetected in the gaps of the AMS active tracking volume or
entering the ECAL from the side.



The following criteria are applied to reject the background events:

\begin{itemize}
\item Identify $p$, He by analyzing the 3-dimensional shower development in ECAL;
\item Identify charged particles by requiring the trajectory direction of the
reconstructed ECAL shower passes inside the AMS sensitive volume and reject these events.
\end{itemize}

The rejection factors for different cosmic ray
species after all cuts have been applied are :  $> 6\times 10^4$ for e$^\pm$, $(2.5\pm 1)\times 10^6$ for $p$ and $> 1.7\times 10^6$  for $He$ nuclei.

\begin{figure}[!ht]
\begin{center}
\includegraphics[clip,width=7.5cm,height=5.6cm]{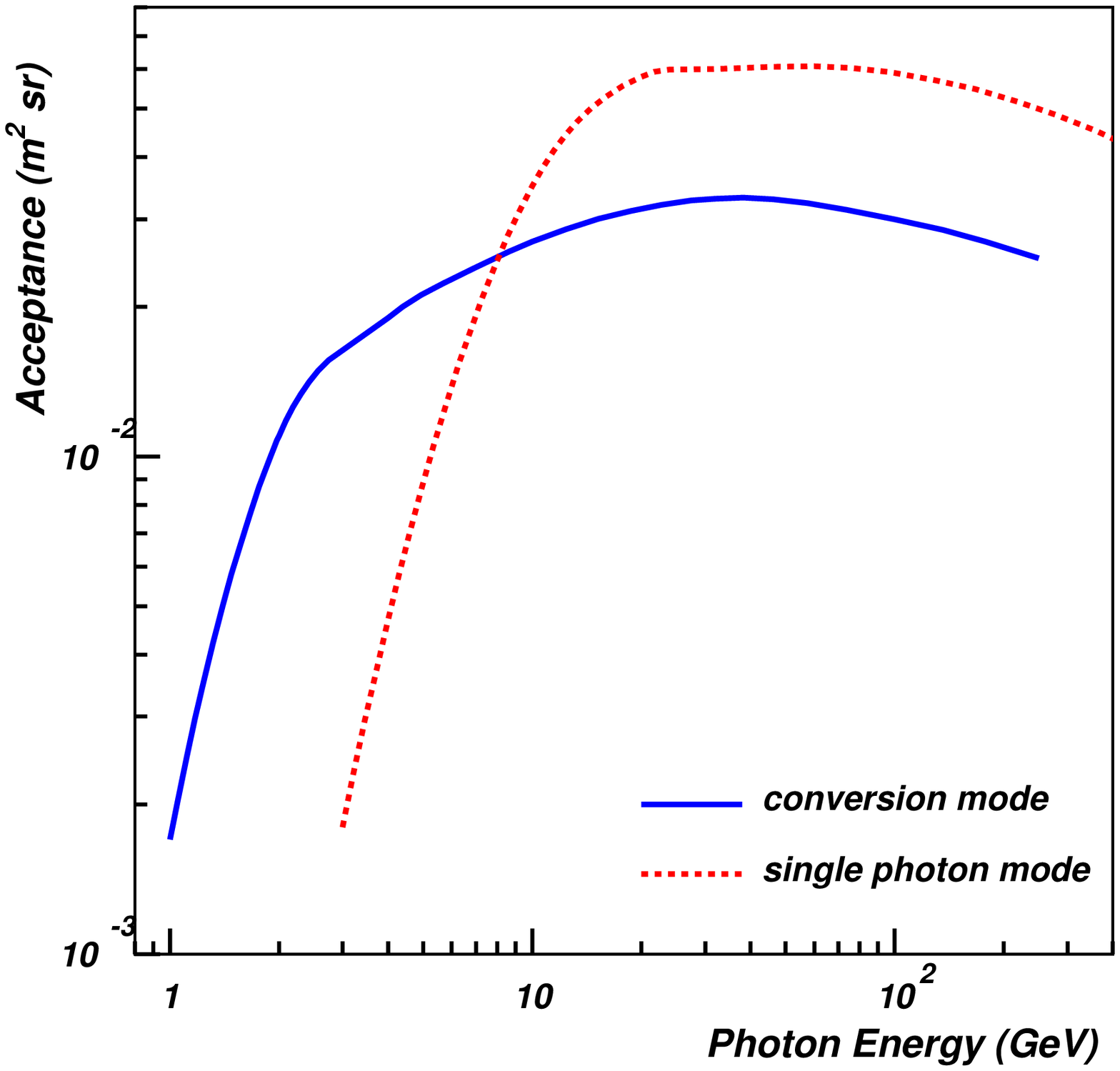}
\includegraphics[clip,width=7.5cm,height=5.6cm]{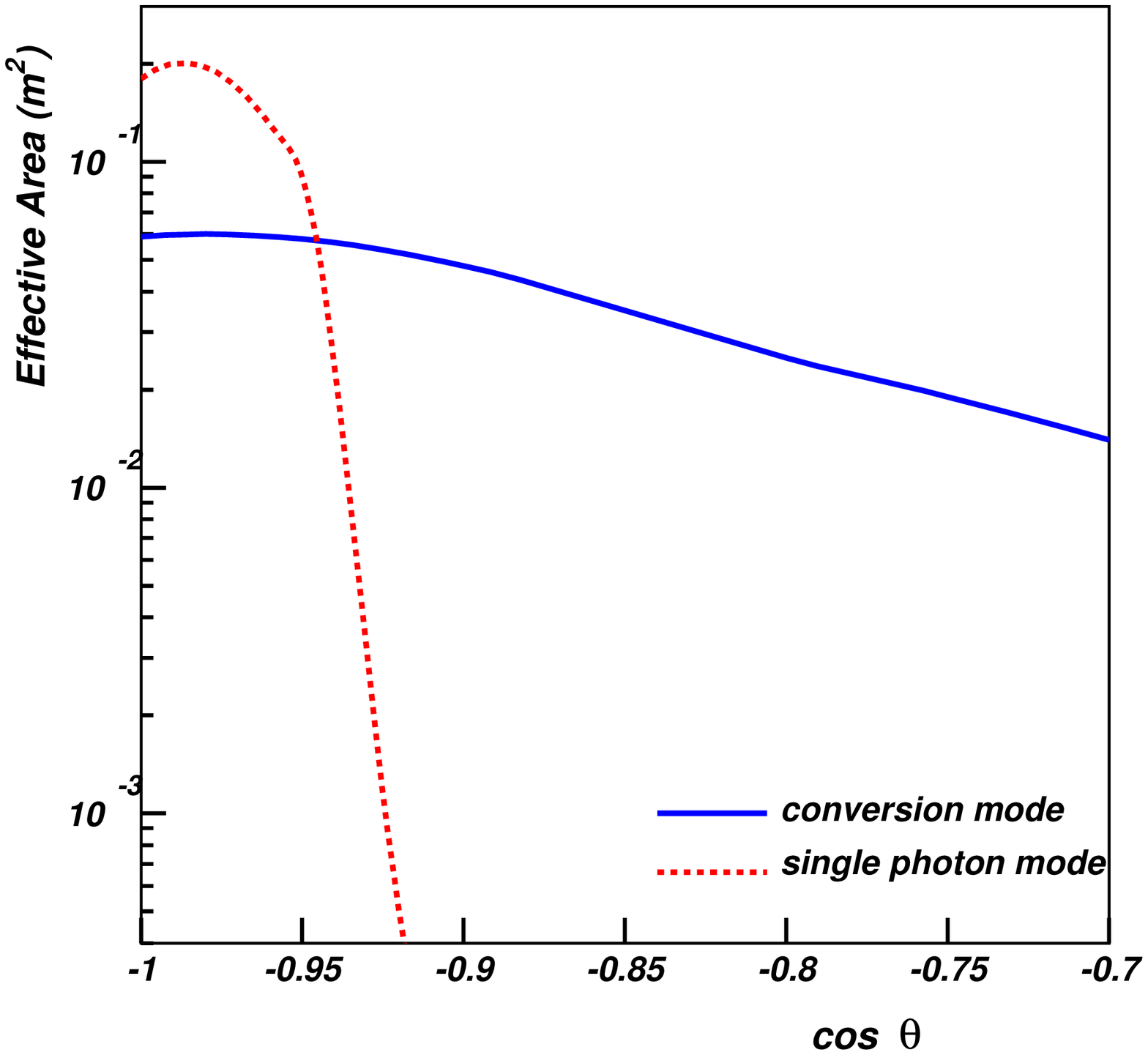}
\end{center}
\caption{AMS acceptance as a function of $\gamma$-ray energy for the two detection modes (top panel). The effective areas versus zenith angle at 50 GeV $\gamma$-ray energy (bottom panel). 
\label{area}}
\end{figure}

\begin{figure}[!hb]
\begin{center}
\vspace*{-10pt}
\includegraphics[clip,width=7.5cm,height=4.5cm]{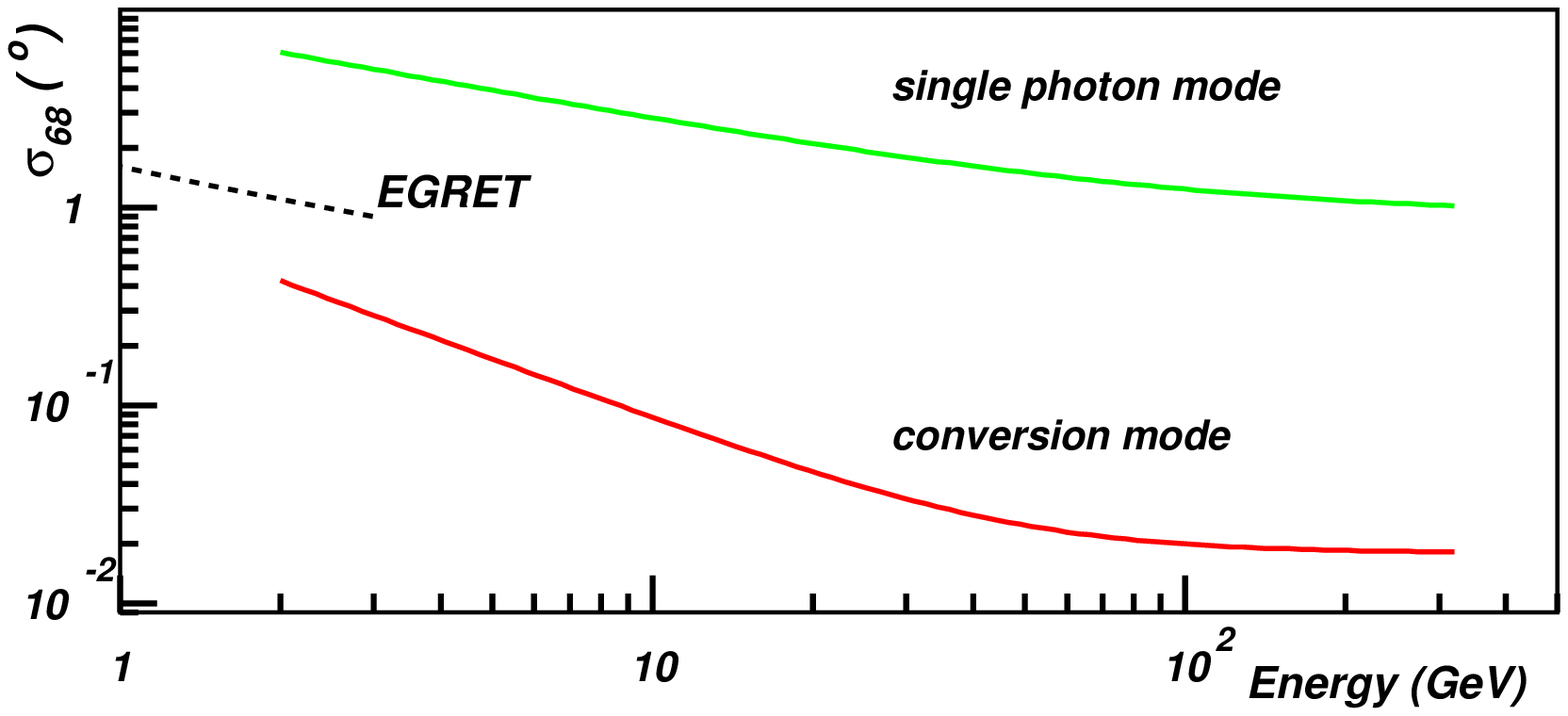}
\includegraphics[clip,width=7.5cm,height=4.5cm]{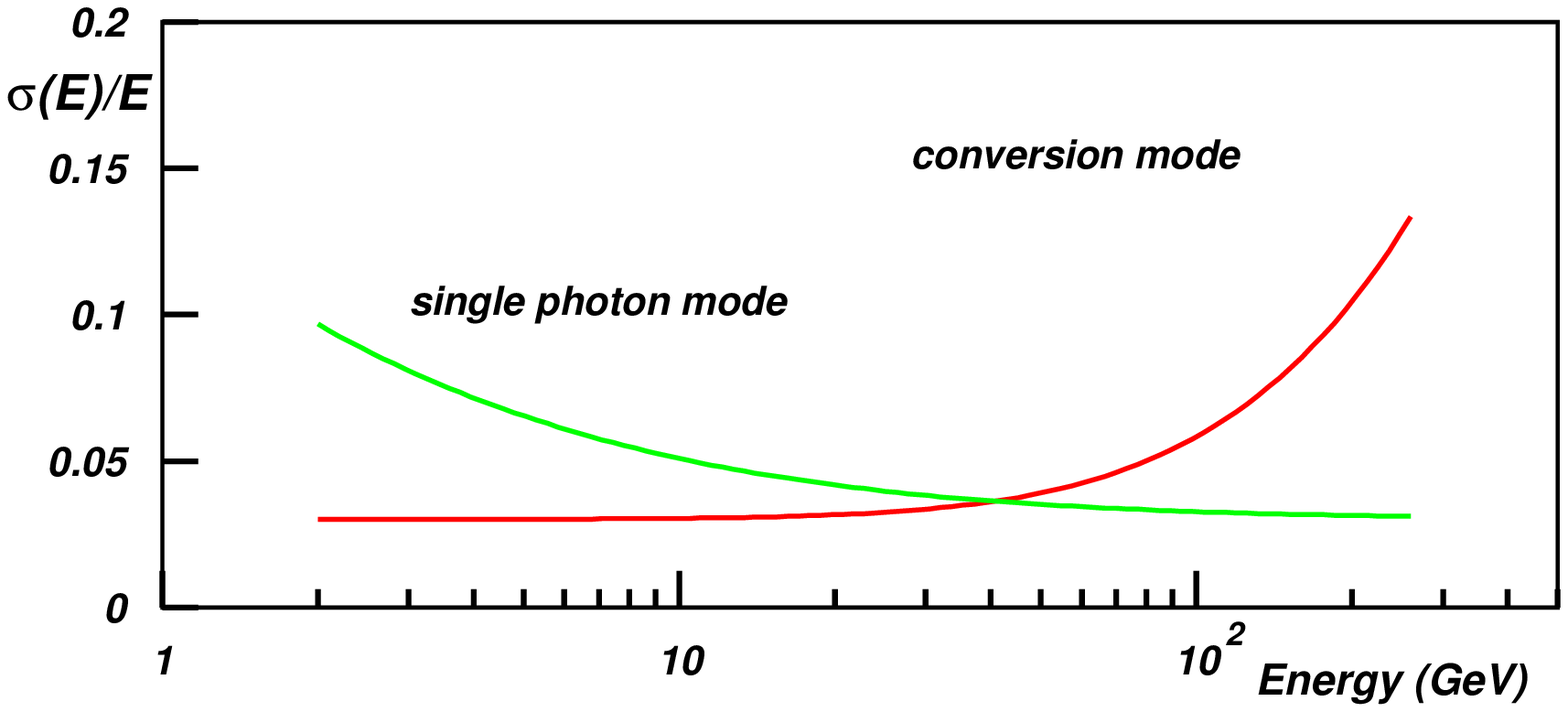}
\end{center}
\caption{The 68\% containement angular resolution for both complementary detection modes as a function of energy (top panel). Energy resolutions as a function of original $\gamma$-ray energy (bottom panel). }
\label{area2}
\end{figure}

\subsubsection{Acceptances and resolutions}
\label{res2}


The simulations are used to parameterize the AMS performance for $\gamma$ detection in terms of acceptance, effective area, angular and energy resolutions, and background rejection.
Figure~\ref{area} shows the acceptance and effective area for the two detection modes. 
The corresponding energy and angular resolutions are shown in Figure~\ref{area2}\footnote{we have chosen a conservative estimate of the single photon mode acceptance; a second study reports a $50\%$ higher acceptance \cite{loic}}. 

The parameterized performance is used to establish the AMS02 sensitivity for the different scenarios in which high energy $\gamma$-rays are produced by the annihilation of dark matter near the Galactic Centre. In a first approximation, we consider a $\gamma$-ray source located at galactic longitude $l$=0 and 
galactic latitude $b$=0. 

\subsection{The sensitivity to the gamma flux and Confidence Level determination}


We have developed a ROOT-based~\cite{root} simulation program, the AMS-$\gamma$ Fast Simulator ($AMSFS$)~\cite{lamannafast}, in order to investigate the AMS capability to localize non-isotropic radiation, either point-like or diffuse. Here we describe the computational approach implemented in the simulator.

We define the detector's $\gamma$-ray source sensitivity as the minimum flux required to achieve a specified 
level of detection significance. The significance $S$ of a detection is given
by:
\begin{equation}
S(E_{\gamma}>E_t) \sim \frac{N^{obs}_{\gamma}(E_{\gamma}>E_t)}{\sqrt{B(E_{\gamma}>E_t)}},
\label{sigma}
\end{equation}
where $N^{obs}_{\gamma}(E_{\gamma}>E_t)$ and $B(E_{\gamma}>E_t)$ are respectively the total number of detected
 photons from the source and the number of background photons falling within
 the source area above an energy threshold $E_t$. $N^{obs}_{\gamma}$ and $B$ are functions of the effective detection 
area $A(E)$ of the instrument, the angular resolution expressed in terms of the 
solid angle $\Omega$$(E)$, the observation time $T_{obs}$ and the differential 
spectra:
\begin{equation}
N^{obs}_{\gamma}(E_{\gamma}>E_t)=\int^{\infty}_{E_t} \int^{ }_{\Omega} \frac{dN_{\gamma}}{dE\;d\Omega} A(E) T_{obs} d\Omega(E)  dE,
\label{sig}
\end{equation}
and
\begin{equation}
B(E_{\gamma}>E_t)=\int^{\infty}_{E_t} \int^{ }_{\Omega} \frac{dB}{dE\;d\Omega} A(E) T_{obs} d\Omega(E) dE.
\end{equation}
In order to establish the significance level of the observation, we require a minimum of three detected gamma events. 

We use the analytical expressions resulting from the best fit to the curves shown in Figures 
\ref{area} and 
\ref{area2} for the energy dependence of the acceptance, the angular and energy resolutions.
The solid angle over which the background is integrated 
for a given source is $\Omega(E)$=$\pi$$\sigma_{68}^2(E)$,
 where $\sigma_{68}$ is the detector angular resolution defined within which 68\% 
of the source photons fall.

The calculation of the Galactic Center observation  time $T_{obs}$ is based on a 3-year misssion on  the International Space Station (ISS)~\cite{orbit}.The AMS observation time is not uniformly distributed over the celestial sphere since the ISS is in a 51.6$^o$ orbit, and the detector is fixed rigidly to the ISS.
Taking into account the precession of the orbital plane of the station about the Earth's pole,
a full sky coverage is obtained about 5.3 times
 per year.
 The $exposure$ (effective area $\times$ the observation time) varies with the photon energy due to the energy dependence the effective area, and the position in the sky, due to the orbit precession. 

The dependence of the effective area on the inclination of the photon direction ($\theta$), the time $dT_{obs}$ spent by the 
detector viewing the Galactic Center within a specific viewing 
inclination $d\theta$, has been calculated and then
integrated over the field-of-view ($\theta$ range up to 42$^o$ for the conversion mode
and 22$^o$ for the single photon mode) and convoluted with the corresponding
effective area: $A(E,d\theta)$ $\times$ $dT_{obs}$. The time intervals when ISS orbits 
over the South Atlantic Anomaly region are excluded.

\noindent
The source spectrum $dN_{\gamma}/dE d\Omega$ in equation (\ref{sig}) corresponds to the photon 
differential spectrum of the dark matter annihilation calculation incorporating the halo 
profile model and the choice of the particle physics parameters, including the mass of the WIMP candidate.
 The background flux $dB/dE d\Omega$ corresponds to the isotropic 
extragalactic $\gamma$-ray background radiation and the galactic 
diffuse radiation (the
latter is due mainly to the decay of $\pi$$^0$s produced by interactions of 
the cosmic rays with
the interstellar medium). The extra-galactic component has been measured by EGRET to 
be~\cite{b11}:
\begin{equation}
\frac{dB_{extragal.}}{dE\;d\Omega}= \Phi_0 \times \left( \frac{E}{k_0}\right)^\epsilon (cm^2\;s\;sr\;GeV)^{-1},
\end{equation}
 where $\Phi$$_0$=(7.32 $\pm$ 0.34) $\times$ 10$^{-6}$ 
(cm$^2$ s sr GeV)$^{-1}$, k$_0$=0.451 GeV and $\epsilon$=-2.10 $\pm$ 0.03.

The galactic diffuse flux is enhanced toward the galactic center and the galactic disk as measured by EGRET. In our calculation we use the parameterization of the differential flux provided in \cite{bergGal}: 

\begin{equation}
\frac{dB_{gal.}}{dE\;d\Omega}= \Gamma_0 \times \left( \frac{E}{r_0}\right)^\alpha (cm^2\;s\;sr\;GeV)^{-1},
\end{equation}
 where $\Gamma$$_0$=8.6 $\times$ 10$^{-5}$ 
(cm$^2$ s sr GeV)$^{-1}$, r$_0$=1 GeV and $\alpha$=-2.7.



Finally, the Galactic center point sensitivity has to take into account the profile of the energy spectrum of the photons produced as a function of the neutralino mass $m_{\chi}$.
For this purpose we define the following function: 

\begin{equation}
\Lambda(m_{\chi}) = \frac{\int^{\infty}_{E_t} \frac{d\Phi}{dE} A(E) dE}{\int^{\infty}_{E_t} \frac{d\Phi}{dE} dE}.
\label{lam}
\end{equation}

\noindent
$\Lambda(m_{\chi})$ includes the
weight of the detector acceptance $A(E)$ on the total rate of photons expected
to be detected for a given differential flux

%
%
\begin{equation}
\frac{d\Phi}{dE} = \frac{1}{4 \pi} \frac{dN_{\gamma}}{dE} \frac{\langle \sigma v \rangle \,}{2\;m_{\chi}^2} \; \Sigma_{19}.
\end{equation}

\begin{figure}[!ht]
\centerline{
\includegraphics[clip,width=8.5cm,height=8.5cm]{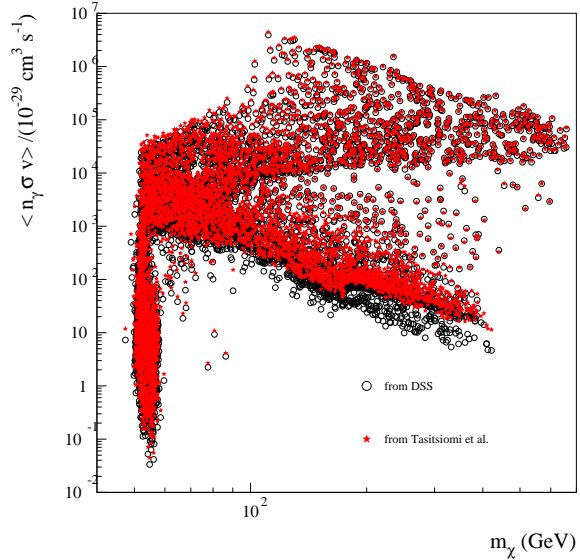}}
\caption{The $\langle N_{\gamma} \sigma v \rangle$ as a function of
$m_{\chi}$ with standard set of parameters as compared with parameterization~\cite{tasi}. A threshold $E_t = 1\,$GeV has been assumed. A normalization factor of 2.4 was applied to obtain compatibility between the two calculations. This factor is used in our calculations.} 
\label{comTaDSS}
\end{figure} 
Therefore we can rewrite the function $\Lambda$ as

\begin{equation}
\Lambda(m_{\chi}) = \frac{\int^{\infty}_{E_t} \frac{dN_{\gamma}}{dE} A(E) dE}{\int^{\infty}_{E_t} \frac{dN_{\gamma}}{dE} dE},
\end{equation}

\noindent
where

\begin{equation}
\int^{\infty}_{E_t} \frac{dN_{\gamma}}{dE} dE = N_{\gamma}(E_{\gamma} \geq E_{t}) 
\end{equation}

\noindent
is the total number of continuum $\gamma$-rays above energy $E_t$ mainly due to the decay of $\pi^0$ mesons produced in jets from neutralino annihilation. For the computation of $N_{\gamma}$ we have considered the parametrization from ref.~\cite{tasi}, after checking its compatibility with PYTHIA parameterization included in DSS as explained in the caption of Figure \ref{comTaDSS}:

\begin{equation}
\begin{split}
N_{\gamma}&(E_{\gamma} \geq E_{t}) =\frac{5}{6} \left( \frac{E_t}{m_{\chi}} \right)^{3/2} - \frac{10}{3}\frac{E_t}{m_{\chi}} +\\
&\quad 5\left(\frac{E_t}{m_{\chi}}\right)^{1/2} + \frac{5}{6}\left(\frac{E_t}{m_{\chi}}\right)^{-1/2} - \frac{10}{3}.
\end{split}
\label{20}
\end{equation}
The differential continuum spectrum, assuming $N_{\gamma} (E_{t} = m_{\chi})$ = 0, is:

\begin{equation}
\begin{split}
\frac{dN_{\gamma}}{dE} &= \frac{1}{m_{\chi}} {\bigg\lgroup}\frac{10}{3} + \frac{5}{12} \left( \frac{E}{m_{\chi}} \right)^{-3/2}-\\
&\quad \frac{5}{4}\left( \frac{E}{m_{\chi}} \right)^{1/2} - \frac{5}{2} \left( \frac{E}{m_{\chi}} \right)^{-1/2} {\bigg\rgroup}.
\end{split}
\label{21}
\end{equation}
With formulas (\ref{20}) and (\ref{21}) we can compute the function $\Lambda(m_{\chi})$ for a given value of the neutralino mass $m_{\chi}$:

\begin{equation}
\Lambda(m_{\chi}) = \frac{1}{N_{\gamma}} \; \int^{\infty}_{E_t} \frac{dN_{\gamma}}{dE} A(E) dE 
\end{equation} 

\noindent
and obtain the corresponding Confidence Level.

The total number of photons detected by AMS is defined as

\begin{equation}
N_{\gamma}^{obs}(E_{\gamma}>E_t) =  T_{obs} \; \int^{\infty}_{E_t} \frac{d\Phi}{dE} A(E) dE.
\end{equation}

\noindent
or, using the $\Lambda(m_{\chi})$ function

\begin{equation}
N_{\gamma}^{obs}(E_{\gamma}>E_t) =  T_{obs} \; \Lambda(m_{\chi}) \; \int^{\infty}_{E_t} \frac{d\Phi}{dE} dE.
\end{equation}
According to the detectability criterion defined in eq.(\ref{sigma}), the minimum detectable flux $F_{min}$, which corresponds to a significance $S(E_{\gamma}>E_t)$=3, is derived by requiring $N\gamma^{obs}$ = $3$ $\sqrt{B}$. The definition of:

\begin{equation}
\Phi_{95} = \int^{\infty}_{E_t} \frac{d\Phi_{min}}{dE} dE   = \frac{3 \; \sqrt{B(E_{\gamma}>E_t)}}{\Lambda(m_{\chi}) T_{obs}}
\end{equation}

\noindent
leads to conservative $95\%-99\%$ Confidence Level values.





\section{Results}

\subsection{mSUGRA and Benchmark point simulations}

The DSS program provides values of the $\gamma$-ray fluxes for the SUSY benchmark
     models \cite{ellis,ellis2}  and the so-called ``wild scan'' configurations
of the mSUGRA parameters.


The SUSY benchmark models have been proposed
to provide a common way of comparing the SUSY discovery potential
of the future accelerators such as LHC or Linear Colliders. The thirteen SUSY
scenarios correspond to different configurations of the five mSUGRA parameters
with the trilinear coupling parameter $A_0$ set to 0.
The models fulfill the conditions imposed by LEP measurements, 
the $g_{\mu} - 2$ result, 
and the relic density constraint of $0.094<\Omega_\chi h^2 < 0.129$.

\noindent
To derive the gamma-ray fluxes for some of
these benchmark models we use our current MC simulation programs: 
DSS, which was described 
previously. In particular 
 the value of $\Omega_\chi h^2$ is calculated in the DarkSUSY part,
while the simultaneous use of the SUSPECT and DarkSUSY package allows to 
perform Renormalization Group Equations evolution from the GUT scale to EWSB scale.

\begin{footnotesize}
\begin{table}
\newcommand{\lstrut}{{$\strut\atop\strut$}}
\caption{The lightest neutralino mass $m_\chi$, the mSUGRA parameters 
$m_0$, $tan{\beta}$, the 
relic neutralino densities, i.e. $\Omega_\chi h^2$ and the values $\langle N_{\gamma} \sigma v \rangle$ as described in the text in units 
of $10^{-29} cm^3\;s^{-1}$; (masses are in GeV and the stars indicate values from~\cite{Gondolo}).}
\label{T:oh2}
\vspace{2mm}
\begin{center}
\begin{tabular}{|c||c|c|c|c|c|}
\hline
model & B & G & I & K & L \\
\hline
&  &  &  & &  \\
$M_{\chi}$ & 98.3 & 153.6 & 143.0 & 571.5 & 187.2 \\
$m_0$  & 59 & 116 &  178 & 999 & 299 \\
$tan{\beta}$ & 10.0 & 20.0 & 35.0 & 38.2 & 47.0 \\
&  &  &  & &  \\
\hline
&  &  &  & &  \\
$\Omega_{\chi} h^2$ & 0.12 & 0.12 & 0.12 & 0.11 & 0.10 \\
&  &  &  & &  \\
$\Omega_{\chi} h^{2*}$ & 0.12 & 0.13 & 0.13 & 0.09 & 0.10 \\
&  &  &  & &  \\
\hline
 &  &  &  & &  \\
$\langle n_{\gamma} \sigma v \rangle$  & 1013 & 1283 & 8380 & 29344 & 33438 \\
 &  &  &  & &  \\
$\langle n_{\gamma} \sigma v \rangle^{*}$  & 782 & 1032 & 6303 & 70903 & 18739 \\
 &  &  &  & &  \\
\hline
\end{tabular}
\end{center}
\end{table}
\end{footnotesize}

Table \ref{T:oh2} presents values of the  lightest neutralino mass 
$m_\chi$ , the mSUGRA parameters $m_0$, 
$tan{\beta}$, the neutralino relic density $\Omega_\chi h^2$ and the values 
$\langle N_{\gamma} \sigma v \rangle$ as described in section 3. 

The corresponding values of neutralino mass, $tan{\beta}$ and $m_0$ are also quoted. A fine-tuning procedure has been applied as in~\cite{ellis} (to fulfill the relic density constraints).
The $N_{\gamma}$ were obtained with fast simulation by the convolution of the differential
$\gamma$-ray fluxes with angular and energy resolution, and applying the acceptance factors of the tracker (TR) and calorimeter (ECAL). 
\begin{footnotesize}
\begin{table}
\newcommand{\lstrut}{{$\strut\atop\strut$}}
\caption{The expected number of photons detected in 3 years for different benchmark models and various dark matter halo profiles. Since the benchmark model flux values are low in the scanned ($\Phi_{\gamma},m_{\chi}$) plane, a 3 GeV energy cut threshold has been applied for signal and diffuse gamma background calculations. The sensitivities ($N_{\sigma}$ = number of standard deviations above diffuse gamma emission) have been calculated only for photons detected in the tracker assuming 3.0 background photons. In the ECAL detector, we expect 196.0 photons for the galactic diffuse gamma emission.}
\label{T:oh1}
\vspace{2mm}
\begin{center}
\begin{tabular}{|c||c|c|c|c|c|}
\hline
model & B & G & I & K & L \\
\hline
 &  &  &  & &  \\
$N_{\gamma}$$^{NFW}_{std}$  & 0.22 & 0.14 & 0.94 & 0.35 & 2.48 \\
 &  &  &  & &  \\
$N_{\gamma}$$^{NFW}_{cuspy}$  & 9.2 & 6.0. & 40.8 & 15.2 & 107.8 \\
 &  &  &  & &  \\
$N_{\gamma}^{Moore}$  & 26.4 & 17.1 & 117.3 & 43.7 & 309.9 \\
 &  &  &  & &  \\
\hline
 &  &  &  & &  \\
$S/B$$^{NFW}_{std}$   & 0.04 & 0.03 & 0.18 & 0.06 & 0.45 \\
 &  &  &  & &  \\
$N_{\sigma}$$^{NFW}_{cuspy}$  & 3.2 & 1.9 & 13.8 & 4.7 & 35.8 \\
 &  &  &  & &  \\
$N_{\sigma}^{Moore}$  & 8.9 & 5.7 & 39.6 & 13.5 & 102.8 \\
 &  &  &  & &  \\
\hline
\end{tabular}
\end{center}
\end{table}
\end{footnotesize}

The choice of
the benchmark model sample was guided by the requirement of meaningful flux values.
These values are low, however, in more favorable astrophysical scenarios, the expected
$N_{\gamma}$ are enhanced by substantial factors varying from 40 in case of
the most cuspy NFW halo profile or about a hundred in case of a Moore profile.

The results in Table \ref{T:oh2} are also compared to those 
in~\cite{Gondolo}, where a
different mSUGRA Monte Carlo was used. 
Good agreement was found between the results of the two calculations.
The observed differences for $\langle N_{\gamma} \sigma v \rangle$, which are at most $\sim$ 25$\%$ between 
our results and those in~\cite{Gondolo}, may be explained by more refined interfacing.

In Table \ref{T:oh1}, $N_{\gamma}$ detected by AMS during 3-year observation and 
the significance values for the benchmark models are presented 
for the NFW halo profile with the standard set of parameter, the most cuspy NFW profile and the Moore profile. 
The diffuse $\gamma$-ray background has been evaluated with the procedure
described in section 3. The hadron contribution can be considered 
negligible for a point-like source, as the proton suppression factors range between
$10^{-5}$ and $10^{-6}$. For the standard NFW profile, only the values of the signal-to-background ratio are given as the expected $N_{\gamma}$ values are not significant.
 

\subsection{Predictions from ``wild scan'' simulations}
 
\subsubsection{mSUGRA results}
We have performed a ``wild scan'' in the mSUGRA parameter space.
Six thousand models have been simulated in the region of
$0.0 \lsim \Omega_{\chi} h^{2} \lsim 0.129$. 
The values below the WMAP lower constraint on $\Omega_{\chi} h^{2}$ (0.094) belong to the additional non-thermal neutralino production scenarios.

\noindent
The ranges of the mSUGRA parameters used in the simulation were:
\begin{center}
sign($\mu$) not constraint\\
$50. < m_{0} < 3000.$\\
$50. < m_{1/2} < 1600.$\\
$0.1 < \vert A_{0} \vert  < 2000.$\\
$3. < tan({\beta)} < 60.$\\
\end{center}

\begin{figure}[!hb]
\centerline{
\includegraphics[clip,width=8.5cm,height=8.5cm]{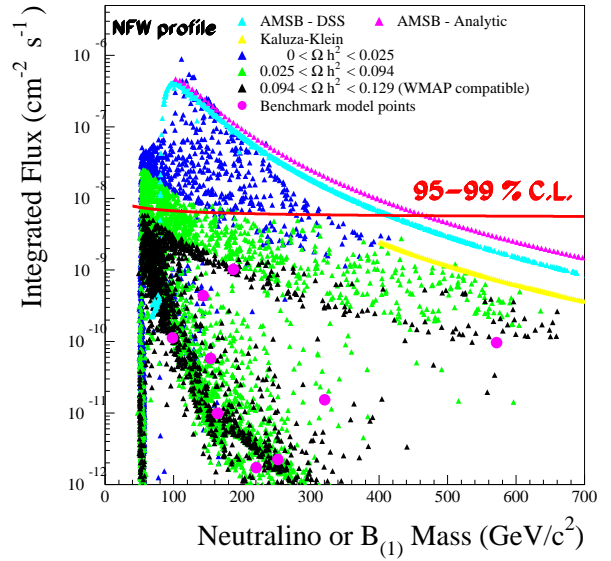}}
\caption{The integrated $\gamma$ flux from the Galactic Centre as a function of
$m_{\chi}$ for the NFW halo profile parameterizations with the standard set of parameters. The considered models are the mSUGRA scheme, AMSB scenario and Kaluza-Klein Universal Extra-dimensions. The various selections were done by varying $\Omega h^2$ cuts.} 
\label{fig5}
\end{figure}

\begin{figure}[!ht]
\centerline{
\includegraphics[clip,width=8.5cm,height=8.5cm]{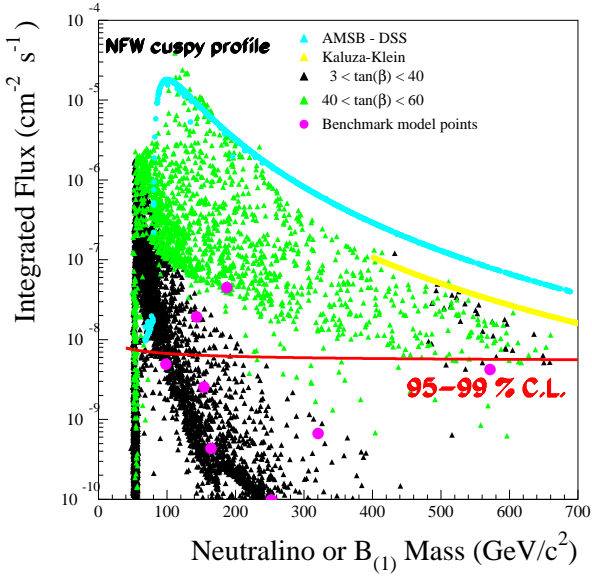}}
\caption{The integrated $\gamma$ flux from the Galactic Centre as a function of
$m_{\chi}$ for a cuspy NFW dark matter halo profile as described in the text.
The considered models are the $mSUGRA$ scheme, $AMSB$ scenario and Kaluza-Klein Universal Extra-dimensions. The various selections were done by varying $tan(\beta)$ cuts.} 
\label{fig6}
\end{figure}

\noindent
The results for the integrated gamma fluxes from the Galactic Centre as a function 
of the $\chi^{0}_{1}$ mass, presented in Figure~\ref{fig5}, 
were obtained for a NFW-$standard$ profile, 
and for a $\gamma$-ray energy threshold of $E_t = 1\,$ GeV. Figure~\ref{fig6} shows the results for the more favorable NFW cuspy dark matter profile.

\subsubsection{AMSB results}
For the prediction of the gamma-ray flux in the $AMSB$ framework, the scheme proposed by SUSPECT was used for the evolution of the AMSB parameters up to the EWSB, as for our mSUGRA simulations. Therefore we use the same DSS interface. 

The constraints set on the four AMSB parameters (as described in~\cite{Gondolo}) were:
\begin{center}
sign($\mu$) not constraint\\
$10^{4} < M_{3/2} < 10^{6}.$\\
$10^{3} < a_{0}  < 15 \cdot 10^{3}$\\
$3. < tan({\beta)} < 60.$\\
\end{center}
For completeness, we have checked that this approach was compatible with the analytical approximation as described below. 

%
From equation (2) we get:
\begin{equation}   
\begin{split}
&\Phi_{\gamma}(\Delta\Omega, E_{\gamma}>E_t)\; = \; 1.19\; 10^{-14}\; N_{\gamma} \left( \frac{\langle \sigma \beta \rangle}{pb}\right) \times\\
&\quad\times \left( \frac{1 TeV}{m_{\chi}} \right)^2\;\Sigma_{19}\; \;\;cm^{-2} s^{-1} 
\end{split}
\label{gr_2}
\end{equation}

\noindent
and, under the hypothesis of a pure Wino LSP, the pair annihilation proceeds by exchange of a charged Wino. The neutral Wino, here assumed to be the WIMP-LSP, can annihilate into a $W$-boson pair ($\tilde{W^0}\tilde{W^0}\rightarrow W^+W^-$). We have considered the results of ~\cite{randal} for the parameterization of the corresponding annihilation cross section, in the non-relativistic limit:

\begin{equation}   
\langle \sigma \beta \rangle \;= \; 9.77 \left( \frac{1 TeV}{m_{\chi}} \right)^2\;\frac{(1-x_W)^{3/2}}{(2-x_W)^{2}} pb 
\label{gr_3}
\end{equation}

\noindent
where $x_W$ = $m_W^2/m_{\chi}^2$.

Thus the equation for the integral flux of photons from Wino LSP annihilation becomes:

\begin{equation}   
\begin{split}
&\Phi_{\gamma}(\Delta\Omega, E_{\gamma}>E_t)\; = \; 1.16\; 10^{-13}\; N_{\gamma} \frac{(1-x_W)^{3/2}}{(2-x_W)^{2}} \times\\
&\quad\times \left( \frac{TeV}{m_{\chi}} \right)^4\;\Sigma_{19}\; \;\;cm^{-2} s^{-1}
\end{split}
\label{gr_4}
\end{equation}

\noindent
where we have used equation (\ref{20}) for the $N_{\gamma}$ value with the scaling factor as described in the caption of Figure~\ref{comTaDSS}.

\begin{figure}[!ht]
\centerline{
\includegraphics[clip,width=8.5cm,height=8.5cm]{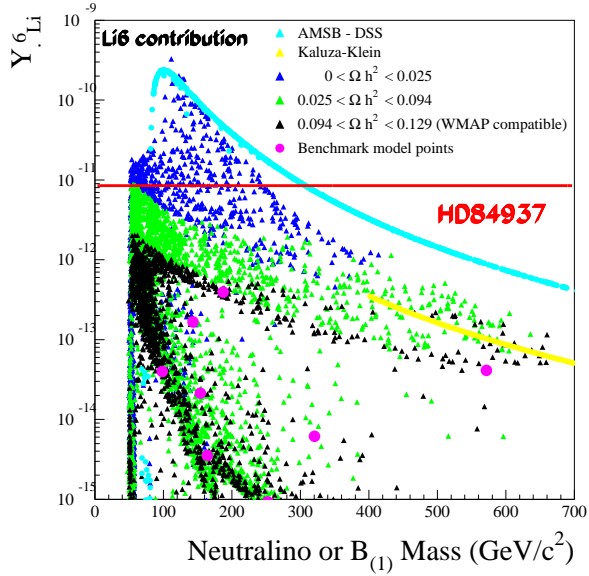}}
\caption{The resulting $^6$Li/H yield synthesized due to residual
neutralino/Kaluza-Klein annihilation during the epoch of Big Bang nucleosynthesis for the same
models as considered before. Also shown is the central value of the $^6$Li/H as observed in the low-metallicity
star HD84937.} 
\label{fig7}
\end{figure}

\subsection{Kaluza Klein results}
The main annihilation channels of LKP into Standard Model particles are charged 
lepton pairs for about 59\% and quark pairs for about 35\%~\cite{bertone}. 
In our case, the calculation of secondary 
gamma ray yield were based on the formulas of
$\sigma$$v$($B^{(1)}$$B^{(1)}$$\rightarrow$$f$$\bar{f}$)~\cite{servant}. The contribution of gamma rays produced from
channels with leptons ~\cite{bergkk2} has been neglected, thus providing more conservative results for
the gamma fluxes at high energies.
As in our previous calculations and following~\cite{bertone} we have used
equation (\ref{gr_2}) with again the $N_{\gamma}$ value from equation (\ref{20}). To obtain the flux for a given halo profile, we have used the corresponding value of $\Sigma_{19}$ given in the paragraph 2.1, thus resulting in:

\begin{equation}   
\begin{split}
&\Phi_{\gamma}(\Delta\Omega, E_{\gamma}>E_t)\; = \; 7.2\; 10^{-15}\; N_{\gamma} \times\\
&\quad\times \left( \frac{1 TeV}{m} \right)^4\;\Sigma_{19}\; \;\;cm^{-2} s^{-1}
\end{split}
\label{kk-s}
\end{equation}
%
%
%

We checked that our results are compatible with~\cite{BertoneRev}.
The expected $\gamma$ fluxes for the AMSB models and Kaluza-Klein models
are also shown in Figures 5 and 6.

\subsection{Sensitivity for considered models}

The $95\%$ CL was obtained by varying, within the uncertainties, the diffuse $\gamma$ background
spectrum as measured by  EGRET \cite{b12} in the Galactic Centre area, 
as described in section 3.3.
With a required minimum of 3 photons over $3\sigma$ of the diffuse gamma
background and a 1 GeV energy threshold
 and 3 year exposure time, the sensitivity to Galactic Center measurements is $(7.0 \pm 0.4) 10^{-9} {\rm cm^{-2} s^{-1}}$ with only a small residual dependence on $m_{\chi}$.
 For both sets of astrophysical conditions, the predicted $\gamma$ fluxes for the 
$AMSB$ and Kaluza-Klein models are above the $95\%$ CL for WIMP masses below 400 GeV. This indicates the
potential of detection or exclusion of AMS02 in the case of the less conventional SUSY scenario
with non-thermal production of neutralinos,
or other dark matter candidate proposed by the Kaluza-Klein extra-dimension theories.


It has been recently shown~\cite{Jeda} that residual dark matter annihilation during the epoch of Big Bang nucleosynthesis may result in an efficient production of $^6$Li. In Figure 7 we show the resulting $^6$Li/H ratio in the dark matter models studied in this paper. The  $^6$Li/H yields have been calculated using the parametrisations given in Ref.~\cite{Jeda}.
 
The predicted abundances are compared to the value reported for the low-metallicity halo stars, such as HD84937, $^6$Li/H $\,\approx 8.47\pm 3.10\times 10^{-12}$~\cite{6Liobs}, (one of the first stars where a $^6$Li detection had been claimed). It is seen that, even in a context of a standard NFW profile (Figure 5), the observed  $^6$Li abundance is consistent with values produced with certain model configurations, in particular in non-thermal scenarios for the $AMSB$ model.
Detections of $^6$Li have been reported for $\sim 10$ other stars~\cite{6Liplateau,nissen}, with abundances comparable to HD84937. 
As this is far from what expected in cosmic ray scenarios which may synthesize $^6$Li, it is possible that the $^6$Li abundance in low-metallicity stars is in fact an indirect signal of dark matter annihilation during Big Bang nucleosynthesis.



\section*{Conclusions}

\noindent
The DarkSUSY and SUSPECT programs were used
to provide the $\gamma$-ray flux predictions from the Galactic Centre region, for the benchmark $mSUGRA$ models, 
the $AMSB$ scenario and Kaluza-Klein Universal Extra-dimensions models, in order to evaluate the discovery potential of AMS for non-baryonic dark matter.
Only models of $mSUGRA$ scenario with 
large $tan\beta$ yield measurable signals 
on a realistic time scale.
This conclusion is confirmed in a second study with a ``wild
scan'' $mSUGRA$ simulations. Various aspects such as dependence
of the results on heavy quark masses or CP-odd Higgs pole contributions
may change the predictions~\cite{M31}.
The most significant signals correspond
to the supersymmetric configurations with the neutralino mass $m_{\chi}$
 $\sim$ 100 GeV.

The sensitivity of the AMS-02 detector for $\gamma$-ray fluxes from a point-like source will allow to detect fluxes smaller by a factor of 2 to 3, compared to those measured by EGRET experiment in the GeV range, in the Galactic Centre region~\cite{b12}. AMS mission will also extend these measuremnets to a poorly explored energy range around 10 GeV, important for the detection of a low mass neutralino. In the TeV range, the Galactic Centre was observed by ground-based Air Shower Cherenkov Telescopes (ACTs), which have detected several intensive astrophysical sources ~\cite{hess,cangaroo}. The signal from the central source, as observed by different ACT experiments, has been analyzed in the context of SUSY dark matter by~\cite{profumo}. More recently, the HESS Collaboration has published a discovery of a diffuse $\gamma$ emission in the galactic plane in TeV range, nearby the SgrA* emitter, possibly produced by the hadronic interactions of the Galactic Cosmic Rays with a complex of molecular clouds~\cite{hess2}. These results indicate that the choice of the observed dark matter source is crucial, and the intense astrophysical environment may be an obstacle for an exotic signal detection. A more promising type of dark matter source could be a nearby Dwarf Spheroidal Galaxy such as DRACO, Sagittarius or Canis Major, presenting lower standard astrophysical backgrounds. These sources have been considered by several authors for the SUSY dark matter flux predictions~\cite{evans,hooper}, and their observation campains have been scheduled by ACT telescope experiments in the coming future.

In the frame of the present study with the Galactic Centre source, the 3 year observation with AMS detector would provide $95\%$ CL exclusion limits
for several $mSUGRA$ models in the case of a favorable dark matter galactic
halo configuration, such as the cuspy or very cuspy NFW profiles. 
Furthermore, if the halo is made of clumps with inner profiles
of cuspy type, or if there is a strong accretion around the central
black hole, the expected signal would increase by two orders of
magnitude. For such cases, given the excellent energy resolution
of the detector, the discovery of a dark matter annihilation signal would
be possible.
In particular, the non-thermal Susy Breaking scenarios, as in case of the $AMSB$ model, result in cosmologically significant $^6$Li abundances, which, when confronted with the results for $^6$Li abundances in low-metallicity stars offer interesting perspectives for indirect dark matter searches and the detection of an annihilation signal by AMS.
We conclude, that a survey of the Galactic Centre by AMS has the potential to contribute significantly to our understanding of dark matter.
 

\section*{Acknowledgements}
We thank all our colleagues from AMS for expressing their support and interest in this study.

\end{document}